\documentstyle[12pt]{jarticle}
\def\si{\quad}
\def\sii{\qquad}
\def\siii{\quad\qquad}
\def\siiii{\qquad\qquad}
\def\vo{\vglue 0.3cm}
\def\voo{\vglue 0.5cm}
\def\vi{\vglue 1cm}

\def\noi{\noindent}

\def\ve{\vfill\eject}
\def\b:{\begin{equation}}
\def\e:{\end{equation}}

\begin{document}

\vglue 1.5cm

\font\twelverm=cmr12
\font\tenrm=cmr10
\font\eightrm=cmr8
\font\twelveit=cmr12
\font\fonta=cmr10 scaled\magstep4
\font\fontb=cmr10 scaled\magstep3
\font\fontc=cmr10 scaled\magstep2
\font\fontd=cmr10 scaled\magstep1


{\Large
\centerline{{\bf Integrable Difference Analogue of the Logistic Equation}}
\centerline{{\bf and B\"acklund Transformation of the KP Hierarchy}}
}
\vi
\centerline{{\bf Noriko SAITOH, Satoru SAITO$^*$ and Akinobu SHIMIZU}}
\voo
\footnotesize{
\centerline{ Department of Applied Mathematics}
\centerline{ Yokohama National University}
\centerline{ 156 Tokiwadai, Hodogaya-ku, Yokohama 240 Japan}

\vo
\centerline{ $^*$ Department of Physics}
\centerline{Tokyo Metropolitan University}
\centerline{ 1-1 Minamiohsawa, Hachiohji, Tokyo 192-03 Japan}
}
\vi
\normalsize
\baselineskip 18pt
\noi

\centerline{
\begin{minipage}[pos]{16cm}
{\bf Abstract}\ -\  A difference analogue of the logistic equation, which
preserves integrability, is derived from Hirota's bilinear difference equation.
The integrability of the map is shown to result from the large symmetry
associated with the B\"acklund transformation of the KP hierarchy. We introduce
a scheme which interpolates between this map and the standard logistic map and
enables us to study integrable and nonintegrable systems on an equal basis. In
particular we study the behavior of Julia set at the point where the
nonintegrable map passes to the integrable map.
\end{minipage}
}

\vi
\centerline{{\bf 1.\ INTRODUCTION}}
\vglue 0.3cm
The completely integrable systems have been revealed to play a fundamental role
in the study of mathematical physics[1,2]. They provide a common basis to
understand phenomena in various fields of mathematical physics. If we try to
deform the systems but preserving their integrability we must generalize the
symmetry possessed by the systems. The quantum group is such an example[3]. The
essential feature of the new symmetry is to replace the differential operators
by their difference analogue[4]. On the other hand it has been known[5] that
many of the completely integrable systems have their discrete counterparts
which are also integrable. A typical example of this sort is the Toda lattice
which is a discrete analogue of the KdV equation along the spatial
direction[6].

It is, however, not obvious that equations obtained by making discretization of
variables of an integrable differential equation are integrable. The KdV
equation can be derived from the Toda lattice by taking an appropriate
continuous limit of the spatial variable. But, if we start from the KdV
equation and discretize the spatial variable we obtain a variety of difference
equations besides the Toda lattice. Those equations obtained by discretizing
the variables in different ways do not preserve their integrability but create
chaos in general. Our main concern in this article is to characterize the
discretization of variables in integrable systems, which maintain their
integrability.

Starting from the (2+1) dimensional Toda lattice and discretizing the two time
variables such that all three variables in the equation appear symmetric, we
arrive at Hirota's bilinear difference equation (HBDE)[7]. This equation is
known completely integrable and is equivalent to the infinite number of soliton
equations belonging to the KP hierarchy[8]. In other words, Hirota's difference
equation is a simultaneous discrete counterpart of soliton equations in the KP
hierarchy. As a matter of fact we can derive all of the soliton equations in
the KP hierarchy from HBDE by taking appropriate continuous limits of the
variables. Therefore HBDE is a desirable object to study for our purpose.

A system which evolves along discrete time can be considered as a dynamical
system if we regard it as an infinite sequence of maps. In this sense our
problem might be thought equivalent to characterize dynamical systems which are
integrable. Since the dependent variable of HBDE is defined on an arbitrary
Riemann surface, we are especially interested in the complex dynamical systems.

{}From physical point of view the existence of discretization of integrable
systems which preserve integrability is important by itself. The intimate
correspondence between the solvable statistical models on two dimensional
lattice and the two dimensional conformal field theories has been noticed from
the begining of the study of the integrable systems[9]. In the relativistic
quantum field theory there arises infinity owing to the fact that the
space-time is a continuous media of the fields. The lattice gauge theory[10]
provides a way to avoid this difficulty. The space-time is replaced by a set of
lattice points. Only at the end of all calculation of physical quantities the
lattice constant is set zero. This approach causes many problems specific to
the way the lattice constant is introduced into the theory. The relativistic
invariance, for instance, is lost in general. If the theory on the lattice is
integrable, however, we are certain that the results obtained on the lattice
are correct in the continuum limit. The systems which preserve integrability
after discretization of variables are also peculiar in the sense that all
results calculated on computer are exact without errors.

In order to clarify the notion of integrability preserving discretization of
integrable systems we like to study, in this paper, the simplest case of
Hirota's bilinear difference equation. We first show that if we consider the
dependence on one variable and ignore the other variables, HBDE reduces to a
difference analogue of the logistic equation, which preserves integrability.
Secondly we argue that the discrete time evolution of this integrable map can
be considered as a special case of the B\"acklund transformation. The existence
of B\"acklund transformation characterizes integrable systems[11]. It enables
one to derive a sequence of new solutions starting from a known solution. Hence
it is quite natural to intepret that the B\"acklund transformation generates a
time evolution of the integrable system. In the last part of this paper we
introduce a scheme which enables us to discuss the completely integrable
systems and nonintegrable systems on the same ground. Specifically we consider
analytic properties of a map which interpolates two different types of
discretization of the logistic equation. Namely we investigate a map which
combines one of the standard logistic map, which presents chaos, and the one
derived from HBDE through the reduction, which is completely integrable.

\ve

\centerline{
{\bf 2.\ DISCRETIZATION OF LOGISTIC EQUATION}
}
\vo
We are interested in characterizing integrable systems which preserve their
integrability after discretization of dependent variables. Most of the
integrable systems discussed in the literature of theoretical physics have
infinite number of freedom and thus too much complicated to be considered as
dynamical systems. Instead of studying them directly we will start our
discussion in this paper from the simplest case, i.e., one dimensional ordinary
differential equations. One of such equations is given by
\b:
{dx\over dt}=f(x).
\e:
This is equivalent to one dimensional Newton equation, which can be solved as
\b:
t=\int^{x(t)}{dx\over f(x)}.
\e:
There exists a unique solution as long as $1/f(x)$ is integrable. If we
discretize the variable $t$ according to Euler's prescription, we obtain,
writing $x(t)\rightarrow z_l$:
\b:
z_{l+h}=z_l+hf(z_l).
\e:
When $h=1$, this equation turns to be an infinite sequence of a map, i.e., a
one dimensional dynamical system.  $h\rightarrow 0 $ reproduces the original
equation. If we consider $z_l$ on the complex plane, and $f(z)$ is a rational
function of $z$ with degree larger than 2, this map is known[12] to have a
Julia set (the closure of repelling periodic points). Therefore this map shows
chaotic behaviour somewhere on the complex plane, and is not integrable. When
$f$ is a polynomial of $z$ of degree 2, the map is called a logistic map and is
studied in detail[12]. In fact if we put
\b:
f(z)=z(\mu -h -\mu z)
\e:
and $h=1$ we get the standard form of the logistic map:
\b:
z_{l+1}=\mu z_l(1-z_l).
\e:
In the limit of $h\rightarrow 0$ it turns to the logistic equation:
\b:
{dx\over dt}=\mu x(1-x).
\e:

The discretization of the logistic equation is not unique. Hence there arises a
question; what is the integrable discretization of the logistic equation? To
answer this question let us recall some properties possessed by integrable
discrete systems. Hirota's bilinear difference equation (HBDE) is one of such
equations whose solutions are completely known[8,11]. It is given by
$$\alpha f(p+1,q,r)f(p,q+1,r+1)+\beta f(p,q+1,r)f(p+1,q,r+1)$$
\b:
\siiii\siiii  +\gamma f(p,q,r+1)f(p+1,q+1,r)=0.
\e:
Here $p,q,r$ are discrete variables and $\alpha,\ \beta,\ \gamma$ are arbitrary
parameters subject to the constraint $\alpha+\beta+\gamma=0$. Infinitely many
equations of the KP hierarchy can be reduced from this single equation by
taking continuous limits of various combinations of the variables and the
parameters[7]. To be more specific we introduce a new set of variables
$\{t_n\}=\{ t_1,t_2,\cdots\}$ by[8]
\b:
t_n={1\over n}\sum_jk_jz_j^n,\sii n\ge 1.
\e:
Here we assume $\sum_jk_j=0$, which is nothing but the energy-momentum
conservation law in the language of the string model[2]. This new set of
variables $\{t_n\}$ are the independent variables of the soliton equations in
the KP hierarchy. The variables $p,q,r$ of HBDE are any three of $k_j$'s of
(8). Therefore we consider infinite number of copies of HBDE. To obtain a
particular soliton equation we change, in one of the HBDE, the variables from
$\{ k_j\}$ to $\{ t_n\}$ and reduce the number of $t$ variables by neglecting
their dependence. From this argument it is apparent that a large symmetry of
the KP hierarchy can be seen if we use the variables $k_j$'s instead of $t_n$'s
of the soliton equations.

The solutions of HBDE are known explicitly and given by analytic functions
defined on Riemann surface with arbitrary genus[13]. $\{z_j\}$ are local
coordinates of the Riemann surface. The space of the solutions is called
universal Grassmannian which is a Grassmann manifold of infinite dimension[14].
The equation which characterizes such a large space of solution is rather
simple if we consider HBDE instead of the set of soliton equations in the KP
hierarchy.

If we wish to discuss integrable systems in comparison with nonintegrable
systems such as logistic map, however, HBDE is still complicated. Therefore we
try to reduce the dependence on the variables. As we simply neglect dependence
on two independent variables, say $q$ and $r$, HBDE becomes trivial. In order
to obtain the most simple nontrivial reduction we introduce new set of
variables
\b:
l=p+q+r+{3\over 2},\si m=-q-{1\over 2},\si n=p+q+1
\e:
and denote by $g$ the dependent variable written by new variables as
$$f(p,q,r)=g_n(l,m).$$
We then obtain
$$\alpha g_n(l+1,m)g_n(l,m+1)+\beta g_n(l,m)g_n(l+1,m+1)\siiii\siii$$
\b:
\siiii\siiii +\gamma g_{n+1}(l+1,m)g_{n-1}(l,m+1)=0.
\e:
The first nontrivial relation is given if we choose
$$
g_1(l,1)=g_0(l,2)=g_2(l,1)\equiv z_l
$$
\b:
g_1(l,2)\equiv c\ ({\rm const.})
\e:
$$
g_n(l,m)=0, \sii {\rm otherwise}
$$
and also
$\beta=-\mu\alpha, \ \gamma=(\mu-1)\alpha,\ c={\mu-1\over\mu}$,
from which follows
\b:
z_{l+1}=\mu z_l(1-z_{l+1}).
\e:
This is quite similar to the standard logistic map (5) except for the second
term on the right hand side. Since this equation is derived by a reduction
starting from an integrable difference equation, the behaviour of solutions
must be completely different from those of the standard map. In fact writing
(12) as
\b:
z_{l+1}={\mu z_l\over 1+\mu z_l}
\e:
the solution is given by
\b:
z_l={\mu^lz_0\over 1+\mu{1-\mu^l\over 1-\mu}z_0}
\e:
for arbitrary initial value $z_0$. This solution is exactly the one which
coincides with one of the logistic equation (6) in the continuous limit. It is
easy to check that this equation itself reduces to the logistic equation in the
same limit. In this sense we call (12) the integrable discretization of the
logistic equation.

The equation (12) itself was studied by Morisita[15] as a difference analogue
of the logistic map and revived by Yamaguti[16] and Hirota[17]. We like to
emphasize here that the integrability of the map (12) has the same origin as
the integrability of the KP hierarchy.

\vglue 1cm
\centerline{
{\bf 3.\ B\"ACKLUND TRANSFORMATION}
}
\vo
As we have seen in the previous section a slight difference of the way of
discretization of an integrable equation leads to a big difference of the
behaviour of the system. The question we ask in this section is how a regular
map is possible and does not fall into chaos. To this end we will show that the
map which maintains regularity is a particular sequence of the B\"acklund
transformations of the KP hierarchy. In the case of the KP hierarchy the
B\"acklund transformations are auto B\"acklund transformations[7,18]. In other
words the transformation generates a sequence of solutions of the same
equation. Starting from a point of the universal Grassmannian a sequence of the
B\"acklund transformations sweeps a trajectory on the space of solutions. From
this result we can glance the deep connection of the integrable dynamical
systems and the large symmetry associated with the B\"acklund transformation.

To begin with we first recall the fact that the time variable $l$ of the map
(13) is a certain combination of $k_j$'s. We like to show that the B\"acklund
transformations of the KP hierarchy are the operations which change values of
$k_j$'s by one. It then follows that there exists a certain transformation
which shifts $l$ into $l+1$ without changing other variables.

The B\"acklund transformation of the KP hierarchy is generated by the
operator[11]
\b:
B(z,w)=\Lambda(z)\bar\Lambda(w)
\e:
where
\b:
\Lambda(z)=\exp\bigl[-\sum^\infty_{n=1}t_nz^{-n}\bigr]\exp\bigl[\sum^\infty_{n=1}{1\over n}z^n\partial_n\bigl],
\e:
\b:
\bar\Lambda(z)=\exp\bigl[\sum^\infty_{n=1}t_nz^{-n}\bigr]\exp\bigl[-\sum^\infty_{n=1}{1\over n}z^n\partial_n\bigl].
\e:
In this expression $\partial_n$ means $\partial /\partial t_n$, where $t_n$'s
are those variables of the soliton equations in the KP hierarchy. Now we
consider the operation of $\Lambda(z_j)$ to $f(k_1,k_2,\cdots)$. The right hand
side factor of $\Lambda(z_j)$ transforms $t_n$ according to
$$
t_n\ \rightarrow\ t_n+{1\over n}z^n_j,\sii n\ge 1.
$$
We observe from the expression of $t_n$ (8) that this change of variables
amounts to replace $k_j$ by $k_j+1$. Similarly the left hand side factor of
$\Lambda(z_j)$ can be rewritten in terms of $k_j$'s. In doing this we must be a
bit careful because $\Lambda(z)$ is not well defined when $z$ approaches to one
of $z_j$'s. Thus we define a new regularized operator by
\b:
\Lambda_q(z_j)\equiv\prod_l(z_j-qz_l)^{k_l}e^{\partial_{k_j}}.
\e:
$\Lambda(z_j)$ is recovered in the $q\rightarrow 1$ limit.
{}From this expression we see that the operation of $B_q(z_i,z_j)\equiv
\Lambda_q(z_i){\bar\Lambda_q}(z_j)$ to one of solutions of HBDE, say
$f(k_1,k_2,\cdots)$, yields
\b:
B_q(z_i,z_j)f(k_1,\cdots,k_i,\cdots,k_j,\cdots)\equiv
b(z_i,z_j)f(k_1,\cdots,k_i+1,\cdots,k_j-1,\cdots),
\e:
where
$$
b(z_i,z_j)=\prod_l\Bigl({z_i-qz_l\over z_j-qz_l}\Bigr)^{k_l}{1\over z_j-qz_i}.
$$
Therefore the operation of $B$ of the B\"acklund transformation (15) causes to
$f(k_1,k_2,\cdots)$ the shifts of the variables $k_i\rightarrow k_i+1,\  k_j
\rightarrow\ k_j-1$. Let $f^{(\alpha)}(k_1,k_2,\cdots)$ be the function
obtained from $f(k_1,k_2,\cdots)$ by the B\"acklund transformation, {\it i.e.,}
\b:
 e^{\alpha B_q(z_i,z_j)}f(k_1,k_2,\cdots)=f^{(\alpha)}(k_1,k_2,\cdots).
\e:
It is well known that the generator of the B\"acklund transformation $B$ has a
fermionic nature, {\it i.e.,} $B^2(z_i,z_j)=0$ holds[11]. Hence we can write
$$\alpha
B_q(z_i,z_j)f(k_1,k_2,\cdots)=f^{(\alpha)}(k_1,k_2,\cdots)-f(k_1,k_2,\cdots).$$

To be more specific let us choose four out of $\{k_j\}$ and call them
$p,q,r,s$. Further we define $l,m,n$ from them according to (9). Then we see
that the increase of $l,m,n$ by one are equivalent to the change of $p,q,r$ as
$$l\rightarrow l+1 \Leftrightarrow r\rightarrow r+1$$
\b:
m\rightarrow m+1 \Leftrightarrow p\rightarrow p+1,\si q\rightarrow q-1
\e:
$$n\rightarrow n+1 \Leftrightarrow p\rightarrow p+1,\si r\rightarrow r-1.$$

If $s$ is the variable associated to the singular point $z_s$ on the Riemann
surface, the amplitude $f$ does not depend on $s$. Letting $z_j$'s associated
to $p,q,r,s$ as $z_p,z_q,z_r,z_s$, the shifts of (21) are realized by the
following operations:
$$l\rightarrow l+1 \Leftrightarrow B_q(z_r,z_s)$$
\b:
m\rightarrow m+1 \Leftrightarrow B_q(z_p,z_q)
\e:
$$n\rightarrow n+1 \Leftrightarrow B_q(z_p,z_r).$$
{}From this we see that under the operation of $B_q(z_r,z_s)$ only the variable
$l$ changes its value whereas other variables remain unchanged. According to
(20) we can express $g_n(l+1,m)$ by $g^{(\alpha)}_n(l,m)$ and $g_n(l,m)$. In
other words $g_n$ at $l+1$ is determined by $g_n^{(\alpha)}$ and $g_n$ at $l$.
Hence it is the generator of the B\"acklund transformation which generates the
time sequence of $z_l$ under the integrable map (13).

Finally we notice about the symmetry corresponding to the B\"acklund
transformation. If we expand $B(z,w)$ into double Laurent series as
\b:
B(z,w)=\sum_{m,n\in {\bf Z}}B_{mn}z^mw^{-n}
\e:
we can show[11]
\b:
[B_{mn},\  B_{m'n'}]=\delta_{m'n}B_{mn'}-\delta_{mn'}B_{m'n}.
\e:
Hence $B$ generates a group $GL(\infty)$.

\vi

\centerline{
{\bf 4.\ AN INTERPOLATION OF INTEGRABLE AND NONINTEGRABLE }
}
\centerline{
{\bf DIFFERENCE ANALOGUES OF THE LOGISTIC EQUATION}
}
\vo
In the previous sections we derived the integrable map associated with logistic
equation starting from HBDE and showed that the integrability of the map owes
to the symmetry associated to the B\"acklund transformation of the KP
hierarchy. We like to clarify in this section how integrable maps can be
characterized among nonintegrable ones. For this purpose we study a map which
interpolates between the integrable and the nonintegrable difference analogues
of the logistic equation, and argue the behaviour of the Julia set.

We have seen in \S 2 that we can derive two different maps (5) and (12)
starting from the same logistic equation (6). Although they look similar at the
level of eqation the behaviour of their solutions are quite different. We like
to know analytically the transition between these two systems. To proceed we
generalize (5) and (12) such that it is a discretization of the logistic
equation (6) and (5) and (12) are included as two particular cases. A simple
generalization is given by
\b:
z_{l+1}=\mu z_l\{ 1-\gamma z_l-(1-\gamma)z_{l+1}\},
\e:
where $\gamma$ is an arbitrary parameter. It is apparent that the standard
logistic map and the integrable map are recovered if $\gamma$ is set 1 and 0
respectively. If we solve this equation for $z_{l+1}$ we find
\b:
z_{l+1}=\phi(z_l),
\e:
where we have introduced a map $\phi(z)$;
\b:
\phi(z)={\mu z(1-\gamma z)\over 1+\mu(1-\gamma)z}.
\e:
This is what we are going to analyze in this section.

We first notice that the map $\phi(z)$ of (27) is similar to the one known as
Blaschke product of degree two[19];
\b:
e^{i\theta}z{\lambda+z\over 1+\bar\lambda z},
\e:
where $\theta$ is an arbitrary real number and $\bar\lambda$ is the complex
conjugate of $\lambda$. This maps the unit circle on itself in the complex
plane. Moreover the Julia set of this map is the unit circle. Since there is no
other unstable fixed point in this map it is very convenient to concentrait our
attention to this to analyze (27) for the first step.

Comparing our map with the expression of the Blaschke product, we see that the
map (27) turns to (28), a Blaschke product, if the parameters satisfy the
relations;
\b:
\mu\gamma=-e^{i\theta}, \ \ \ \mu={-1\over \bar\gamma(1-\gamma)}.
\e:
We are interested in the case of small values of $|\gamma|$. As far as (29) are
satisfied, the Julia set remains on the unit circle. This happens only if the
value of $\mu$ diverges when $\gamma$ approaches to zero. Hence we can not draw
any conclusion for the limit of $\gamma\rightarrow 0$, since the value of $\mu$
in our interest is finite.

In order to proceed further we try to see the map from different coordinate
system in the complex plane. Namely we introduce a new variable $w$ which is a
M\"obius transformation of $z$, i.e., $w=\psi(z)$. In this new coordinate
system, $\phi$ turns to $\tilde\phi$ following to
\b:
\phi(z)\ \rightarrow\ \tilde\phi(z)=\psi^{-1}\circ\phi\circ\psi(z).
\e:
The map $\phi(z)$ of (27) has three fixed points at $0,\infty$ and
$1-{1\over\mu}$. We want 0 and $\infty$ to remain as fixed points after the
change of the variable, so that the new map still keeps the form
\b:
\tilde\phi(z)=e^{i\theta}z{\lambda+z\over 1+\lambda' z}.
\e:
One of such coordinate transformations is
\b:
\psi(z)={1-{1\over\mu}\over(\gamma+\mu-\mu\gamma)e^{i\theta}z+1},
\e:
which yields a new map of the form (31) with
\b:
\lambda={1+\gamma-\mu\gamma\over\mu+\gamma-\mu\gamma}e^{-i\theta},\si \lambda'
=\mu e^{i\theta}.
\e:
In the new coordinate the fixed points are at 0, $\infty$ and
$$p=-e^{-i\theta}{1\over \mu+\gamma-\mu\gamma}.$$
The multipliers of the fixed points at $0,\ \infty\  {\rm and}\  p$ are given,
respectively, as
\b:
\lambda_0={1+\gamma-\mu\gamma\over\mu+\gamma-\mu\gamma},\si
\lambda_\infty=\mu,\si \lambda_p=1-{1\over \gamma}.
\e:
Notice that $z=p$ is a repelling fixed point irrespective to the value of $\mu$
when $|\gamma|$ is small. The nature of the fixed point at $\infty$ is uniquely
determined by the value of $\mu$ alone.

If $\mu$ varies independent on the value of $\gamma$, the product
$\lambda_0\cdot \lambda_\infty$ goes to 1 as $\gamma$ tends to 0. This means
that, for small values of $|\gamma|$, the origin is attractive when $\infty$ is
repelling and vice versa. This situation can be also understood from the map
itself. Letting $\gamma$ be zero in (31) we get
\b:
\tilde\phi(z)={1\over \mu}z.
\e:
The sequence of this map attracts all points in the complex plane to the origin
(resp. $\infty$) if $|\mu|$ is larger (resp. smaller) than 1. Therefore the
Julia set reduces to a point at $\infty$ or 0 depending on the value of $|\mu|$
is larger or smaller than 1, respectively.

The condition for the map (31) to be a Blaschke product is
\b:
{1+\gamma-\mu\gamma\over\mu+\gamma-\mu\gamma}=\bar\mu,\si{\rm or}\si
\gamma={|\mu|^2-1\over |\mu-1|^2}.
\e:
Under the constraint (36) for the parameters we calculate the values of the
multipliers (34). We then find $\lambda_0=\bar\mu, \  \lambda_\infty=\mu$,
hence the origin as well as the $\infty$ are attractive fixed points when
$|\mu|<1$. The third fixed point is at
$$p=e^{-i\theta}{\bar\mu -1\over \mu -1}.$$
Therefore $p$ is on the unit circle and is a repelling point where $\gamma$
takes  a negative real value. It then follows that all points inside of the
unit circle are attracted to the origin by the map whereas the points outside
of the circle are pushed away to infinity. The behaviour of the map on the unit
circle is chaotic, because the Julia set is the unit circle. It can not escape
from the unit circle as long as $\mu$ satisfies the constraint (36) for a given
value of $\gamma$. This is true also when the value of $|\gamma|$ approaches to
zero. Does this mean that the Julia set remains at $\gamma=0$ ? To see this
point more carefully, we restrict the value of $\mu$ to satisfy (36) and write
the map explicitly;
\b:
\tilde\phi(z)=e^{i\theta}z{\bar\mu e^{-i\theta}+z\over 1+\mu e^{i\theta}z}\ =\
\mu^{-1}z{\bar\mu + e^{i\theta}z\over \mu^{-1}+e^{i\theta} z}.
\e:

Now let $\gamma$ go to zero along the negative real axis, so that $|\mu|<1$  is
satisfied. The absolute value of $\mu$ tends to one owing to the constraint
(36). The multipliers at 0 and $\infty$ approach to 1, and thus these points
cease to be attractors. Then the map is given by (35) again as we see from
(37). In this case, however, $\mu$ is pure phase and the sequence of this map
forces every point on the complex plane to rotate uniformly around the origin.
The angle of the rotation is determined by the phase of $\mu$. From (36) we can
calculate the value of the phase. Writing $\mu=|\mu|e^{i2\pi\chi}$ it turns out
to be
$$2\pi\chi=\lim_{\gamma\rightarrow 0}\arccos\Bigl( 1+{1-|\mu|\over
\gamma}\Bigr).$$
Therefore the manner of the rotation is uniquely determined by the value of the
limit of the ratio of $1-|\mu|$ and $\gamma$.

It is known that a sequence of the rotation of an irrational number of phase
causes one dimensional ergodic behaviour of zero entropy[20]. If we choose the
ratio between $1-|\mu|$ and $\gamma$ arbitrarily the phase of $\mu$ will fall
to an irrational number with the probability one as $\gamma$ approaches to
zero. Then the rotation of all $z$ on the complex plane caused by (37) are most
probably ergodic with entropy 0, except for the neutral fixed points at 0 and
$\infty$. The Julia set, which was on the unit circle when $\gamma<0$, does not
exist when $|\gamma|$ is set equal zero.

Let us summarize the above results. As the value of $\gamma$ approaches to 0 we
have three cases:

1) if $|\mu| > 1$, the Julia set is at $\infty$,

2) if $|\mu| < 1$, the Julia set is at 0,

3) if $|\mu| = 1$, there exists no Julia set but all the points between 0 and
$\infty$ on the complex plane rotate ergodically so long as the ratio between
$1-|\mu|$ and $\gamma$ does not go to a rational number as $\gamma$ goes to 0.

We can further proceed the analysis of our map (27) from the mathematical point
of view. As a complex dynamical system the map is interesting by itself. Since
further investigation requires more preparation, however, we shall report it
elsewhere[21].

\vi
\centerline{
{\bf \si DISCUSSIONS}
}
\vo
The completely integrable systems constitute a small portion exceptional among
general nonlinear systems and behave quite different from others. Nevertheless
they describe important nature of fundamental problems in various fields of
theoretical physics. The large symmetry of the integrable systems enables us to
determine their solutions analytically, whereas the missing of some conserved
currents prevents us to predict the behaviour of solutions in the generic case.
The soliton type of equations are completely integrable systems whose detail
features have been studied in the last two decades. In order to proceed further
and to understand the real importance of the role played by integrable systems
in theoretical physics it is necessary to study them in comparison with
nonintegrable systems. In other words the real understanding of the basic
problems in theoretical physics will be achieved when the complete integrable
systems are fully characterized among general nonlinear systems.

There has been, however, no systematic way to investigate nonintegrable
systems. The nonintegrable systems which have been studied are mainly one
dimensional dynamical systems. Even the simplest one of such systems is far
from totally investigated. This should be contrasted with the case of the
completely integrable systems, in which the most interesting systems are
infinite dimensional and solutions are given analytically. Therefore it is
difficult to find a way to discuss these two different systems on a common
basis.

In this paper we have shown that the simplest reduction of Hirota's bilinear
difference equation turns out to be an integrable difference analogue of the
logistic equation. We provided a scheme which interpolates this equation and
the standard logistic map. It enables us to study integrable and nonintegrable
maps on the equal ground. In particular we concentrated our attention to the
behaviour of the Julia set at the point where the system turns from
nonintegrable to integrable.

The integrable part we studied here, is a small portion of HBDE. It will be the
most desirable if there is a simple scheme in which HBDE itself is included as
an integrable part of some nonlinear system. Although we have not found such a
scheme we could interpret the integrability of the map results from the large
symmetry possessed by the integrable system such as $GL(\infty)$ which
generates the B\"acklund transformation of the KP hierarchy.

\ve

\centerline{{\bf REFERENCES}}
\vo
\begin{enumerate}
\footnotesize
\item
 See, for example, ``Infinite Analysis'', {\it ed.} A.Tsuchiya, T.Eguchi and
M.Jimbo (World Scientific, Singapore, 1992); ``Yang-Baxter Equation in
Integrable
Systems'', {\it ed.} M.Jimbo (World Scientific, Singapore, 1990).

\item S.Saito, Phys. Rev. {\bf D36} (1987) 1819; Phys. Rev. Lett. {\bf 59}
(1987) 1798; Phys. Rev. {\bf D37} (1988) 990; H.C$\hat {\rm a}$teau and
S.Saito, Phys. Rev. Lett. {\bf 65} (1990) 2487; in ``Strings '88'' {\it ed.}
S.J.Gates,Jr., C.R.Preitschopf and W.Siegel (World Scientific, 1989) p.436 ; in
``Nonlinear Fields; Classical, Random, Semiclassical'', {\it ed.}
P.Garbaczewski and Z.Popowicz, (World Scientific,1991) p.268.

\item M.Jimbo, Lett. Math. Phys. {\bf 10}, (1985) 63; {\bf 11}, (1986) 247;
Int. Jour. Mod. Phys. {\bf A4} (1989) 3759; V.G.Drin'feld, Sov. Math. Dokl.
{\bf 32} (1985) 254; ``Quantum Groups'', ICM Proceedings, (Berkeley, 1986)
P.898.

\item I.B.Frenkel and N.Yu.Reshetikhin, Comm. Math. Phys. {\bf 146} (1992) 1.

\item E.Date, M.Jimbo and T.Miwa, J. Phys. Soc. Jpn. {\bf 51} (1982) 4116,
4125.

\item M.Toda, J. Phys. Soc. Jpn. {\bf 22} (1967) 431; ``Theory of Nonlinear
Lattices'' (Springer-Verlag, 1981); N.Saitoh, J. Phys. Soc. Jpn. {\bf 49}
(1980) 409.

\item R.Hirota, J. Phys. Soc. Jpn. {\bf 50} (1981) 3787; ``Nonlinear Integrable
Systems'' {\it ed.} M.Jimbo and T.Miwa (World Scientific, 1983) p.17.

\item T. Miwa, Proc. Jpn. Acad. {\bf 58A} (1982) 9.

\item A.Belavin, A.Polyakov and A.Zamolodchikov, Nucl. Phys. {\bf B241} (1984)
333.

\item See, for example, C.Rebbi, ``Lattice Gauge Theories abd Monte Carlo
Simulasions'', (World Scientific, Singapore 1893).

\item E.Date, M.Jimbo, M.Kashiwara and T.Miwa, J. Phys. Soc. Jpn, {\bf 50}
(1980) 3806, 3813; M.Jimbo and T.Miwa, Publ. RIMS. Kyoto Univ. {\bf 19} (1983)
943.

\item See, for example, R.L.Devaney, "An Introduction to Chaotic Dynamical

Systems", 2nd ed. (Addison-Wesley, 1989).

\item I.M.Krichever, Russian Math. Surveys {\bf 32} (1977) 185;

M.Mulase, J. Diff. Geom. {\bf 19} (1984) 403.

\item M.Sato and Y.Sato, Publ. RIMS. (Kyoto Univ.) {\bf 388} (1980) 183 ;

{\it ibid.} {\bf 414} (1981) 181; M.Sato, {\it ibid.} {\bf 433} (1981) 30.

\item M.Morisita, Res. Popul. Ecol. {\bf VII} (1965) 52.

\item M.Yamaguti, Mathematics on Nonlinear Phenomena (in Japanese)

(Asakura-Shoten, Tokyo 1972).

\item R.Hirota, Applied Mathematics {\bf 3} (in Japanese) (1993) 48.

\item S.Saito and N.Saitoh, J. Math. Phys. {\bf 28} (1987) 1052; Phys. Let.
{\bf A 123} (1987) 283; N.Saitoh and S.Saito, J. Phys. Soc. Jpn {\bf 56} (1987)
1664; J. Phys. A:Math. Gen. {\bf 23} (1990) 3017.

\item See, for example, J.Milnor, ``Dynamics in One Complex Variable:
Introductory Lectures'' SUNY preprint \#1990/5 (1990).

\item P.Billingsley, ``Ergodic Theory and Information'' (John Wiley and Sons.
Inc., New York, 1965).

\item N.Saitoh, S.Saito and A.Shimizu, ``An Analysis of a Family of Rational
Maps which Contains Integrable and Nonintegrable Difference Analogue of the
Logistic Equation'', in preparation.
\end{enumerate}

\end{document}